\documentclass[a4paper]{article}

\usepackage{INTERSPEECH2020}
\usepackage{amsmath}
\usepackage{multirow}
\usepackage{comment}

\newcommand{\norm}[1]{\left\lVert#1\right\rVert}

\title{Adversarial Transformation of Spoofing Attacks for Voice Biometrics}
\name{Alejandro Gomez-Alanis, Jose A. Gonzalez-Lopez, Antonio M. Peinado\thanks{This work has been supported by the Spanish Ministry of Science and Innovation Project No. PID2019-104206GB-I00/AEI/10.13039/501100011033. Alejandro Gomez-Alanis holds a FPU fellowship from the Spanish Ministry of Education (FPU16/05490). Jose A. Gonzalez-Lopez holds a Juan de la Cierva-Incorporación fellowship from the Spanish Ministry of Science, Innovation and Universities (IJCI-2017-32926). We also acknowledge the support of NVIDIA Corporation with the donation of a Titan X GPU.}}

\address{
  University of Granada
}
\email{agomezalanis@ugr.es, joseangl@ugr.es, amp@ugr.es}

\begin{document}

\maketitle
%
\begin{abstract}

Voice biometric systems based on automatic speaker verification (ASV) are exposed to \textit{spoofing} attacks which may compromise their security. To increase the robustness against such attacks, anti-spoofing or presentation attack detection (PAD) systems have been proposed for the detection of replay, synthesis and voice conversion based attacks. Recently, the scientific community has shown that PAD systems are also vulnerable to adversarial attacks. However, to the best of our knowledge, no previous work have studied the robustness of full voice biometrics systems (ASV + PAD) to these new types of adversarial \textit{spoofing} attacks. In this work, we develop a new adversarial biometrics transformation network (ABTN) which jointly processes the loss of the PAD and ASV systems in order to generate white-box and black-box adversarial \textit{spoofing} attacks. The core idea of this system is to generate adversarial \textit{spoofing} attacks which are able to fool the PAD system without being detected by the ASV system. The experiments were carried out on the ASVspoof 2019 corpus, including both logical access (LA) and physical access (PA) scenarios. The experimental results show that the proposed ABTN clearly outperforms some well-known adversarial techniques in both white-box and black-box attack scenarios. 


  
\end{abstract}

\section{Introduction}
 
Voice biometrics aims to authenticate the identity claimed by a given individual based on the speech samples measured from his/her voice. Automatic speaker verification (ASV) \cite{asv} is the conventional way to put voice biometrics into practical usage. However, in recent years, ASV technology has been shown to be at risk of security threats performed by impostors who want to gain fraudulent access by presenting speech resembling the voice of a legitimate user \cite{kde-access, lcgrnn-taslp}. Impostors could use either logical access (LA) attacks \cite{ASVSpoof2015}, such as text-to-speech synthesis (TTS) and voice conversion (VC) based attacks, or physical access (PA) attacks such as replay based attacks \cite{ASVSpoof2017}.

\begin{figure}[!t]
\centering
\includegraphics[width=0.5\textwidth]{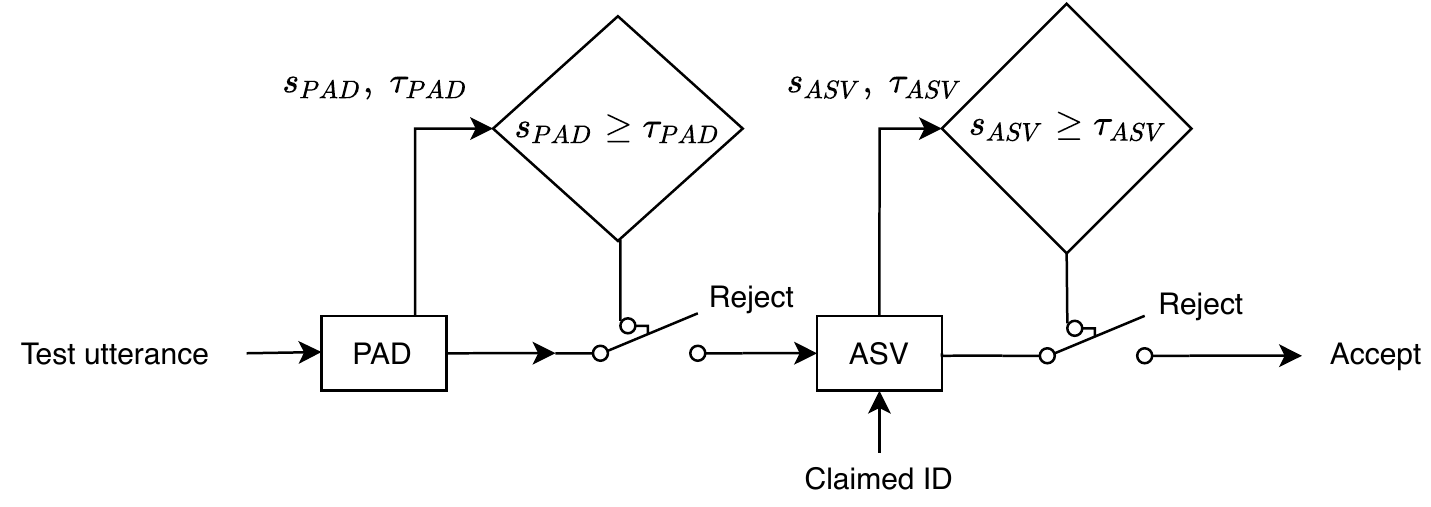}
\caption{Block diagram of a score-level cascaded integration biometrics system. $s_{PAD}$, $\tau_{PAD}$ and $s_{ASV}$, $\tau_{ASV}$ denote the scores and thresholds of the PAD and ASV systems, respectively.}
\label{fig:biometrics_system}
\end{figure}

To protect voice biometrics systems \cite{tifs}, it is common to develop anti-spoofing or presentation attack detection (PAD) \cite{pad_nomenclature} techniques which allow for differentiating between \textit{bonafide} and \textit{spoofing} speech \cite{Mine, iberspeech18, lcgrnn-interspeech}. Typically, the resulting biometrics system is a score-level cascaded integration of PAD and ASV subsystems, as depicted in Fig.~\ref{fig:biometrics_system}. This is the same integration as the one used in the last two ASVspoof challenges \cite{ASVSpoof2017, ASVspoof2019}. 

To make things more complex, different investigations \cite{asru_adv, icassp_adv} have recently shown that PAD systems are also vulnerable to adversarial attacks \cite{adversarial_attacks}. These attacks can easily fool deep neural network (DNN) models by perturbing benign samples in a way normally imperceptible to humans \cite{adv_2013}. Adversarial attacks can be divided into two main categories: white-box and black-box attacks. In this work, we refer to white-box attacks as those where the attacker can access all the information of the victim model (i.e., model architecture and its weights). Likewise, we will use the term black-box for those attacks where the attacker does not know any information about the victim model but it can be queried multiple times in order to estimate a surrogate model (student) of the victim model (teacher), using the binary responses (acceptance/rejection) of the victim model as ground-truth labels.

The main contributions of this work are:

\begin{itemize}

\item Investigate the robustness of full voice biometrics systems (ASV + PAD) under the presence of adversarial \textit{spoofing} attacks.

\item Propose an adversarial biometrics transformation network (ABTN) which is able to generate adversarial \textit{spoofing} attacks in order to fool the PAD system without being detected by the ASV system.

\item To the best of our knowledge, adversarial \textit{spoofing} attacks have only been studied on logical access scenarios (TTS and VC based attacks). In this work, we also include physical access scenarios (replay based attacks).

\end{itemize}

The rest of this paper is organized as follows. Section \ref{sec:related_work} outlines some well-known adversarial attacks employed as baselines in this work. Then, in Section \ref{sec:method}, we describe the proposed ABTN for white-box and black-box scenarios. After that, Section \ref{sec:setup} outlines the speech corpora, systems details, and metrics employed in the experiments. Section \ref{sec:results} discusses the experimental results. Finally, we summarize the conclusions derived from this research in Section \ref{sec:conclusion}.

\section{Background}
\label{sec:related_work}



Adversarial \textit{spoofing} examples can be generated by adding a minimally perceptible perturbation to the input \textit{spoofing} utterance in order to do a refinement of the \textit{spoofing} attack. 
In this work, we focus on targeted attacks, which aim to fool the PAD system by maximizing the probability of a targeted class (\textit{bonafide}) different from the correct class (\textit{spoof}). Specifically, to generate adversarial \textit{spoofing} attacks, we fix the parameters $\boldsymbol{\theta}$ of a well-trained DNN-based PAD model and perform gradient descent to update the \textit{spoofing} spectra of the input utterance so that the PAD model classifies it as a \textit{bonafide} utterance. Mathematically, our goal is to find a sufficiently small perturbation $\boldsymbol{\delta}$ which satisfies:

\begin{equation}
\begin{split}
& \boldsymbol{\tilde{X}} = \boldsymbol{X} + \boldsymbol{\delta}, \\
& f_{\boldsymbol{\theta}}(\boldsymbol{X}) = y, \\
& f_{\boldsymbol{\theta}}(\boldsymbol{\tilde{X}}) = \tilde{y},
\end{split}
\end{equation}

\noindent where $f$ is a well-trained DNN-based PAD model parameterized by $\boldsymbol{\theta}$, $\boldsymbol{X}$ denotes the sequence of speech feature vectors extracted from the input \textit{spoofing} utterance (short time Fourier transform (STFT), typically), $y$ is the true label corresponding to $\boldsymbol{X}$, $\tilde{y}$ is the targeted label class of the attack (\textit{bonafide} class), $\boldsymbol{\tilde{X}}$ denotes the perturbed input features, and $\boldsymbol{\delta}$ is the additive perturbation. Typically, $\boldsymbol{\Delta}$ is the feasible set of the allowed perturbation $\boldsymbol{\delta}$ ($\boldsymbol{\delta} \in \boldsymbol{\Delta}$), which formalizes the manipulative power of the adversarial attack. Normally, $\boldsymbol{\Delta}$ is a small $l_{\infty}$-norm ball, that is, $\boldsymbol{\Delta} = \{ \boldsymbol{\delta} \mid \norm{\boldsymbol{\delta}}_{\infty} \leq \epsilon \}$, $\epsilon \geq 0 \in \mathbb{R}$.

There are multiple ways to generate the perturbation $\boldsymbol{\delta}$, where the fast gradient sign method (FGSM) \cite{fgsm} and the projected gradient descent (PGD) \cite{pgd} methods are the most popular adversarial attack procedures. The FGSM attack consists of taking a single step along the direction of the gradient, i.e., 

\begin{equation}
    \boldsymbol{\delta} = \epsilon \cdot \text{sign}(\nabla_{\boldsymbol{X}} Loss(\boldsymbol{\theta}, \boldsymbol{X}, y)),
\end{equation}

\noindent where $Loss$ denotes the loss function of the neural network ($\boldsymbol{\theta}$), and the $\text{sign}$ method simply takes the sign of its gradient. Unlike the FGSM, which is a single-step method, the PGD is an iterative method. Starting from the original input utterance $\boldsymbol{X}_0 = \boldsymbol{X}$, the input utterance is iteratively updated as follows:

\begin{equation}
\begin{split}
& \boldsymbol{X}_{n+1} = \text{clip}(\boldsymbol{X}_{n} + \alpha \cdot \text{sign}(\nabla_{\boldsymbol{X}} Loss(\boldsymbol{\theta}, \boldsymbol{X}, y)), \\
& \text{for} \; n = 0, ..., N-1,
\end{split}
\end{equation}

\noindent where $n=0, ..., N-1$ is the iteration index, $N$ is the number of iterations, $\alpha = \epsilon / N$, and the $\text{clip}()$ function applies element-wise clipping such that $\norm{\boldsymbol{X}_{n} - \boldsymbol{X}}_{\infty} \leq \epsilon$, $\epsilon \geq 0 \in \mathbb{R}$.






\section{Proposed method}
\label{sec:method}

The performance of the FGSM and PGD methods is limited by the possibility of sticking at local optima of the loss function. Moreover, both methods have a limited search space ($\boldsymbol{\Delta}$) so that the perturbed \textit{spoofing} speech $\boldsymbol{\tilde{X}}$ is perceptually indistinguishable from the original \textit{spoofing} speech $\boldsymbol{X}$.

In this work, we propose the Adversarial Biometrics Transformation Network (ABTN), which is a neural network that transforms a \textit{spoofing} speech signal into an adversarial \textit{spoofing} speech signal against a target biometrics system. Formally, an ABTN can be defined as a neural network $g_{f,h}: \boldsymbol{X} \rightarrow \boldsymbol{\tilde{X}}$, where $f(\boldsymbol{X})$ and $h(\boldsymbol{X})$ are the PAD and ASV models of the target biometrics system, respectively. The PAD and ASV models can provide either a probability distribution across class labels (white-box scenario) or just a binary decision (black-box scenario). In both scenarios, the objective of the ABTN is to generate adversarial \textit{spoofing} attacks from \textit{spoofing} speech in order to fool the PAD system while not being detected by the ASV system, i.e., while not modifying the speaker information.



\subsection{White-box scenario}

\begin{figure}[t]
\centering
\hspace*{-0.8cm}\includegraphics[width=0.55\textwidth]{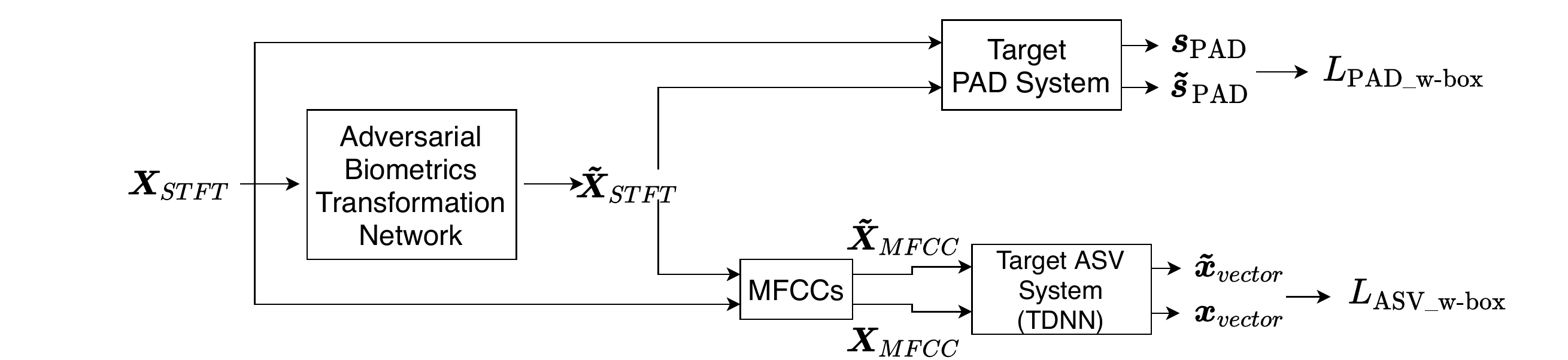}
\caption{Proposed adversarial biometrics transformation system for white-box scenarios.}
\label{fig:atn_whitebox}
\end{figure}

The architecture of the proposed ABTN system for the white-box scenario is depicted in Fig.~\ref{fig:atn_whitebox}. The output of the ABTN is fed into the target biometrics system which is composed of a PAD and an ASV system based on a time-delay neural network (TDNN) \cite{xvectors} for x-vector extraction (in fact, this is the only component of the ASV system that we need). The objective of this system is to train the ABTN so that it can generate adversarial attacks from \textit{spoofing} speech which are able to fool the PAD system while, at the same time, it does not cause any changes to the ASV output (i.e., it does not change the speaker representation given by the corresponding x-vector). To train the ABTN, the PAD and ASV network parameters are frozen but the gradients are computed along them in order to back-propagate them to the ABTN parameters. To find the optimal parameters of the ABTN in the white-box (w-box) scenario, we minimize the following loss function:

\begin{equation}
    L_{\text{w-box}} = L_{\text{PAD\_w-box}}(\boldsymbol{s}_{\text{PAD}},\boldsymbol{\tilde{s}}_{\text{PAD}}) + \beta \cdot L_{\text{ASV\_w-box}}(\boldsymbol{x}_{\text{vector}}, \boldsymbol{\tilde{x}}_{\text{vector}}),
\end{equation}

\noindent where,

\begin{equation}
    L_{\text{PAD\_w-box}}(\boldsymbol{s}_{\text{PAD}},\boldsymbol{\tilde{s}}_{\text{PAD}}) = \norm{r_{\alpha}(\boldsymbol{s}_{\text{PAD}}) - \boldsymbol{\tilde{s}}_{\text{PAD}}}_{2},
\end{equation}

\begin{equation}
    L_{\text{ASV\_w-box}}(\boldsymbol{x}_{\text{vector}}, \boldsymbol{\tilde{x}}_{\text{vector}}) = \norm{\boldsymbol{x}_{\text{vector}} - \boldsymbol{\tilde{x}}_{\text{vector}}}_{2}.
\end{equation}

\noindent $L_{\text{PAD\_w-box}}$ and $L_{\text{ASV\_w-box}}$ are the loss components associated to the PAD and ASV systems, respectively, and $\beta$ is a hyper-parameter to weight the importance of the two losses. $\boldsymbol{s}_{\text{PAD}}$ and $\boldsymbol{\tilde{s}}_{\text{PAD}}$ are the probability output vectors from the PAD system of the original and adversarial \textit{spoofing} utterances, respectively. Likewise, $\boldsymbol{x}_{\text{vector}}$ and $\boldsymbol{\tilde{x}}_{\text{vector}}$ denote the x-vectors of the original and adversarial  \textit{spoofing} utterances, respectively, and $r_{\alpha}$ is a reranking function which can be formulated as

\begin{equation}
     r_{\alpha}(\boldsymbol{s}_{\text{PAD}}) = norm \Bigg(
        \begin{cases} 
          \alpha \cdot \text{max}(\boldsymbol{s}_{\text{PAD}}) & k = 0 \\
          \boldsymbol{s}_{\text{PAD}}(k) & k \neq 0 
       \end{cases} \Bigg),
\end{equation}

\noindent where $k$ is the index class variable of the $\boldsymbol{s}_{\text{PAD}}$ probability vector,  $\alpha > 1$ is an additional hyper-parameter which defines how large $\boldsymbol{s}_{\text{PAD}}(k=0)$, i.e., the probability of the \textit{bonafide} class, is with respect to the current maximum probability class, and $norm$ is a normalizing function which rescales its input to be a valid probability distribution.

\subsection{Black-box scenario}

\begin{figure}[t]
\centering
\hspace*{-1.0cm}\includegraphics[width=0.56\textwidth]{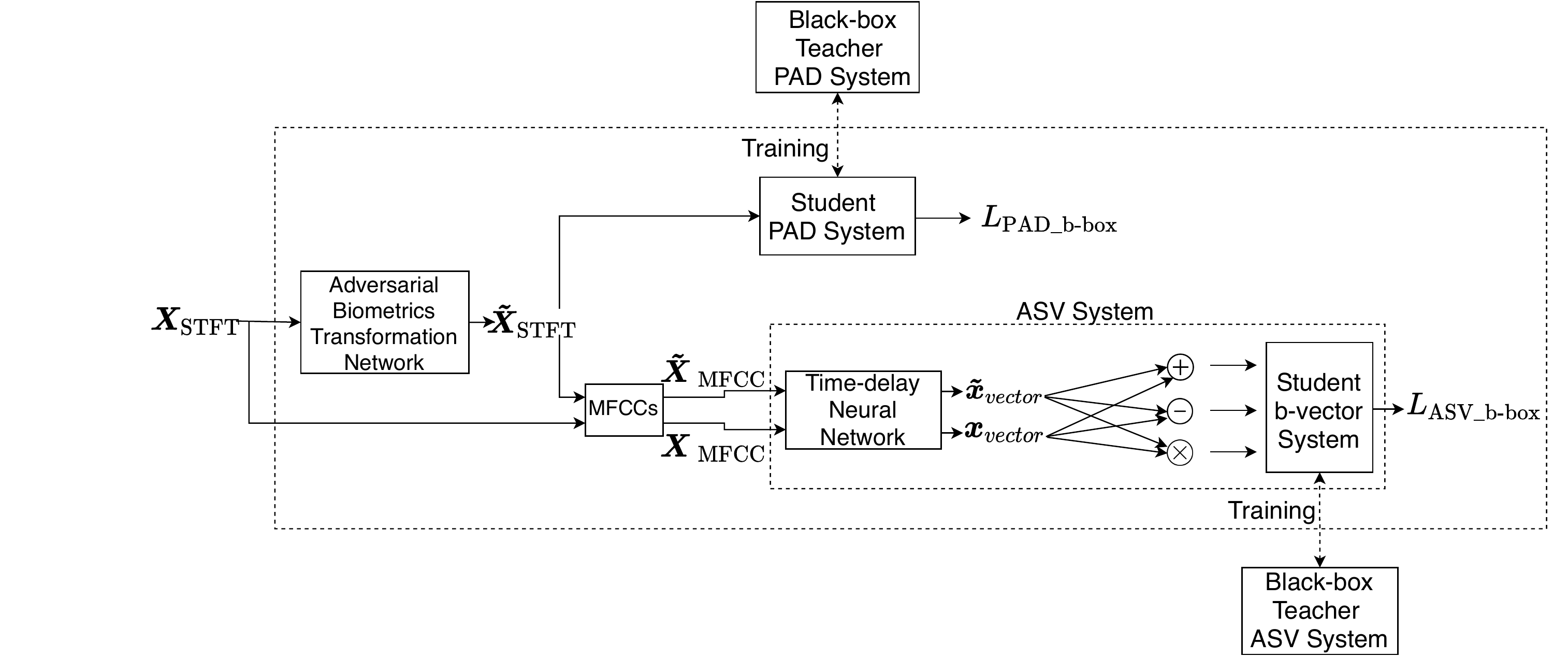}
\caption{Proposed adversarial biometrics transformation system for black-box scenarios.}
\label{fig:atn_blackbox}
\end{figure}

The architecture of the proposed ABTN system for the black-box scenario is depicted in Fig.~\ref{fig:atn_blackbox}. 
Similarly to the white-box scenario, the objective of this system is to generate adversarial attacks from \textit{spoofing} speech which are able to fool the target (teacher) PAD system and, at the same time, bypass the target (teacher) ASV system by not modifying the speaker information represented by the corresponding x-vector. However, the limitation of the black-box scenario is that we do not have access to the parameters of the target biometrics system. Thus, we train a student PAD and a b-vector \cite{b-vector} based ASV systems by making requests to the target black-box biometrics system which only responds with a binary decision of acceptance or rejection, using these binary decisions as ground-truth labels. Therefore, the student PAD and b-vector systems are trained as binary classifiers in order to mimick the performance of the teacher PAD and ASV systems, respectively. Specifically, the student b-vector system computes the probability that the two input x-vectors belong to the same speaker, i.e., that $P(b(\boldsymbol{x}_{\text{vector}},\boldsymbol{\tilde{x}}_{\text{vector}}) = 1)$, where $b$ denotes the b-vector model.

To train the ABTN in the black-box scenario, the student PAD and ASV network parameters are also frozen but the gradients are computed along them in order to back-propagate them to the ABTN parameters. To find the optimal parameters of the ABTN in the black-box (b-box) scenario, we minimize the following loss function:

\begin{equation}
    L_{\text{b-box}} =  L_{\text{PAD\_b-box}}(\boldsymbol{\tilde{s}}_{\text{PAD}}) + \beta \cdot L_{\text{ASV\_b-box}}(\boldsymbol{x}_{\text{vector}}, \boldsymbol{\tilde{x}}_{\text{vector}}),
\end{equation}

\noindent where,

\begin{equation}
    L_{\text{PAD\_b-box}}(\boldsymbol{\tilde{s}}_{\text{PAD}}) = \norm{onehot(k=0) - \boldsymbol{\tilde{s}}_{\text{PAD}}}_{2},
\end{equation}

\begin{equation}
    L_{\text{ASV\_b-box}}(\boldsymbol{x}_{\text{vector}}, \boldsymbol{\tilde{x}}_{\text{vector}}) = 1 - P(b(\boldsymbol{x}_{\text{vector}},\boldsymbol{\tilde{x}}_{\text{vector}}) = 1).
\end{equation}

\noindent $L_{\text{PAD\_b-box}}$ and $L_{\text{ASV\_b-box}}$ are the loss components associated to the PAD and ASV systems, respectively. Moreover, the function $onehot$ denotes the one-hot function and $k=0$ is the index of the \textit{bonafide} class, so that the PAD system is fooled by firing the input \textit{spoofing} utterance as a \textit{bonafide} utterance.

\section{Experimental Setup}
\label{sec:setup}

This section briefly describes the speech corpora and metrics employed in our experiments, as well as the details of the proposed system.

\subsection{Speech corpora}

We conducted experiments on the ASVspoof 2019 database \cite{ASVspoof2019-full} which is split into two partitions for the assessment of LA and PA scenarios. This database also includes protocols for evaluating the performance of PAD, ASV and integration (biometric) systems. Thus, we used this corpus for training the standalone PAD systems in the LA and PA scenarios, separately. Then, we generated adversarial \textit{spoofing} attacks using only the \textit{spoofing} utterances, so that they can bypass the biometric system. We did not generate any adversarial examples from \textit{bonafide} utterances, since we argue that they would not be \textit{bonafide} anymore.

On the other hand, we also employed the Voxceleb1 \cite{voxceleb} to train a TDNN \cite{xvectors} as an x-vector extractor for the ASV system. Also, following \cite{tifs}, a b-vector \cite{b-vector} ASV scoring system was trained in the black-box scenario using the \textit{bonafide} utterances from the ASVspoof 2019 and Voxceleb1 development datasets.

\subsection{Spectral analysis}

Speech signals were analyzed using a Hanning analysis windows of 25 ms length with 10 ms of frame shift. Log-power magnitude spectrum features (STFT) with 256 frequency bins were obtained to feed all the PAD systems. The ASV systems were fed with Mel-frequency cepstral coefficients (MFCCs) obtained with the Kaldi recipe \cite{kaldi_recipe}. Only the first 600 frames of each utterance were used to extract acoustic features.

\subsection{Implementation details}

Two state-of-the-art PAD systems were adapted from different works, i.e., a light convolutional neural network (LCNN) \cite{kde-access} and a Squeeze-Excitation network (SENet50) \cite{assert}. The PAD scores were directly obtained from the \textit{bonafide} class of the softmax output. For ASV, a TDNN x-vector model \cite{xvectors} was trained as an embedding extractor. Then, a probabilistic linear discriminant analysis (PLDA) \cite{plda} and a b-vector system \cite{b-vector} were trained as ASV scoring systems.

The proposed ABTN is formed by five convolutional layers with 16, 32, 48, 48 and 3 channels, respectively, and a kernel size of $3 \times 3$, followed by leaky ReLU activations. It was trained using the Adam optimizer \cite{Adam} with a learning rate of $3 \cdot 10^{-4}$. Also, early stopping was applied to stop the training process when no improvement of the loss across the validation set was obtained. The values of $\alpha$ and $\beta$  were empirically set to 10 and 0.001, respectively, using a grid search on the validation set. 

\begin{table*}[!t]
\renewcommand{\arraystretch}{1.3}
\vspace*{-0.3cm}
\centering
\vspace*{-0.25cm}
\setlength{\tabcolsep}{4.8pt}
	\begin{tabular}{| c | c | c | c | c || c | c | c | c |}
	\hline
	\multirow{2}{*}{System} &
      \multicolumn{4}{c||}{Logical Access Attacks} &
      \multicolumn{4}{c|}{Physical Access Attacks} \\
    \cline{2-9}
    & $\text{EER}_{\text{spoof}}$(\%) & $\text{EER}_{\text{ASV}}$(\%) & $\text{EER}_{\text{joint}}$(\%) & min-tDCF & $\text{EER}_{\text{spoof}}$(\%) & $\text{EER}_{\text{ASV}}$(\%)  & $\text{EER}_{\text{joint}}$(\%) & min-tDCF  \\
	\hline
	\hline
	No Attack & 5.91 & 31.10* & 20.13 & 0.1252 & 4.77 & 18.62* & 13.37 & 0.1238 \\
	\hline
	FGSM ($\epsilon = 0.1$) & 5.98 & 31.14* & 20.32 & 0.1279 & 7.50 & 18.65* & 15.47 & 0.2157 \\
	PGD ($\epsilon = 0.1$) & 5.95 & 31.13* & 20.25 & 0.1267 & 6.08 & 18.63* & 14.38 & 0.1717 \\
	FGSM ($\epsilon = 1.0$) & 8.15 & 31.53* & 25.44 & 0.1287 & 35.64 & 18.71* & 26.54 & 0.9335 \\
	PGD ($\epsilon = 1.0$) & 7.02 & 31.46* & 25.37 & 0.1266 & 44.42 & 18.83* & 26.77 & 0.9665 \\
	FGSM ($\epsilon = 2.0$) & 2.01 & 30.11* & 14.13 & 0.0623 & 1.02 & 17.61* & 11.82 & 0.0380 \\
    PGD ($\epsilon = 2.0$) & 4.97  & 31.38* & 22.62 & 0.1078 & 29.29 & 18.44* & 25.28 & 0.8677 \\
	FGSM ($\epsilon = 5.0$) & 0.00 & 19.46* & 2.45 & 0.0000 & 0.00 & 11.37* & 11.79 & 0.0000 \\
	PGD ($\epsilon = 5.0$) & 0.16 & 19.09* & 2.56 & 0.0058 & 0.00 & 9.48* & 11.79 & 0.0000 \\
	Proposed ABTN & \textbf{35.19} & \textbf{31.52}* & \textbf{39.15} & \textbf{0.5829} & \textbf{95.17} & \textbf{18.87}* & \textbf{36.63} & \textbf{1.0000} \\
\hline
\end{tabular}
\caption{\label{table:asv_eval_set} Results of the black-box adversarial attacks on the ASVspoof 2019 logical access (LA) and physical access (PA) test sets in terms of $\text{EER}_{\text{spoof}}$(\%), $\text{EER}_{\text{ASV}}$(\%), $\text{EER}_{\text{joint}}$(\%) and min-tDCF. The target PAD system is based on a LCNN, while the student PAD system is based on a SENet50. The target ASV system is based on a TDNN + PLDA, while the student ASV system is based on a TDNN + b-vector. (*) The ASV evaluation includes both \textit{bonafide} and \textit{spoofing} utterances. 
}
\vspace*{-0.3cm}
\end{table*}

\subsection{Evaluation setup}

The PAD systems were evaluated using the pooled equal error rate ($\text{EER}_{\text{spoof}}$) across all attacks. Likewise, the ASV systems were also evaluated using the $\text{EER}_{\text{ASV}}$, employing both \textit{bonafide} utterances (target and non-target) and \textit{spoofing} utterances. Any utterance rejected by either the PAD or ASV subsystems was assigned arbitrarily a $-\infty$ score for computing the integration performance. Then, the integration (biometrics) systems were evaluated using the joint EER ($\text{EER}_{\text{joint}}$) and the minimum normalized detection cost function (min-tDCF) \cite{min_tdcf} with the same configuration as the one employed in the ASVspoof 2019 challenge \cite{ASVspoof2019}. All the PAD, ASV and biometrics systems were evaluated using the ASVspoof 2019 test datasets.

\section{Experimental Results}
\label{sec:results}

The performance of the baseline biometrics system is shown in Table \ref{table:asv_eval_set} as 'No Attack'. The LA and PA PAD systems are among the best single systems evaluated in the ASVspoof 2019 challenge \cite{ASVspoof2019}. The ASV system yields an EER of 4.75 and 7.25\% in the LA and PA datasets when evaluating only the target and non-target \textit{bonafide} utterances. However, its performance is degraded to 31.10 and 18.62\% in the LA and PA test datasets when the \textit{spoofing} utterances are also evaluated, as shown in Table \ref{table:asv_eval_set}.

\subsection{White-box scenario}

\begin{figure}[!t]
\centering
\includegraphics[width=0.5\textwidth]{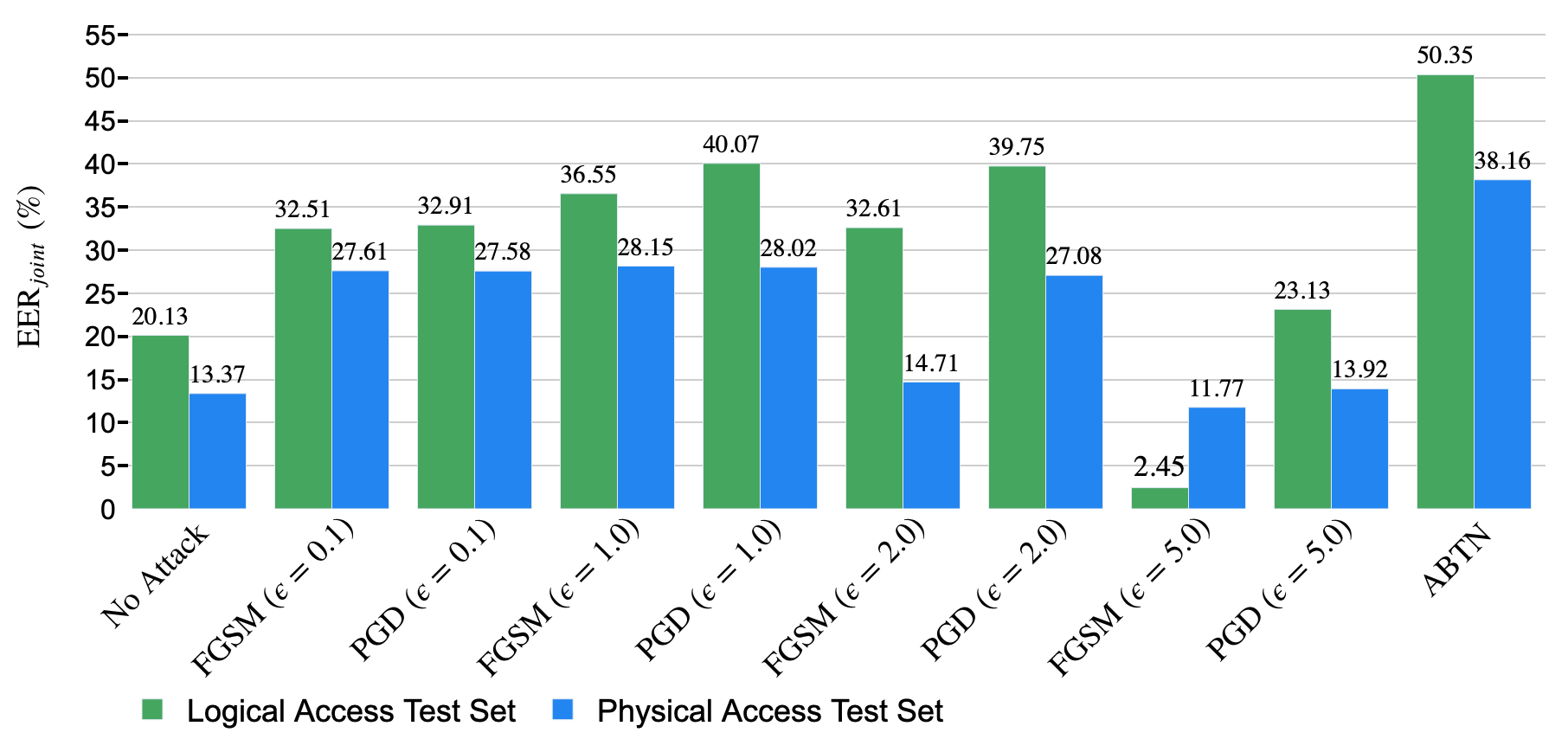}
\caption{$\text{EER}_{\text{joint}}$(\%) of the white-box adversarial attacks on the ASVspoof 2019 logical and physical access test sets.}
\label{fig:whitebox_results}
\end{figure}

Fig.~\ref{fig:whitebox_results} shows the $\text{EER}_{\text{joint}}$ of the white-box adversarial attacks evaluated in the ASVspoof 2019 LA and PA test sets. The PAD and ASV systems are the state-of-the-art LCNN and TDNN + PLDA, respectively. As it was expected, PGD achieves slightly better results than FGSM due to its iterative procedure for generating the adversarial attacks. Moreover, the proposed ABTN outperforms the rest of adversarial attacks, obtaining 10.28\% and 10.14\% higher $\text{EER}_{\text{joint}}$ with respect to the best PGD configuration ($\epsilon=1.0$) in the LA and PA test sets, respectively. It is worth noticing that when the hyper-parameter $\epsilon$ of the FGSM and PGD methods is equal or higher than $2.0$, the biometrics system is able to detect the perturbation noise added by these adversarial attacks. In these cases, the performance of the \textit{spoofing} attacks is even worse than when not using any adversarial attack (denoted by 'No Attack').

\subsection{Black-box scenario}

Table~\ref{table:asv_eval_set} shows the performance metrics for the black-box scenario. The target biometrics system consists of the same state-of-the-art LCNN (PAD) and TDNN + PLDA (ASV) systems evaluated in the previous section. The student PAD and ASV systems are the SENet50 and the TDNN + b-vector systems, respectively. 

The proposed ABTN attacks outperform the best FGSM and PGD configurations by 27.04 and 50.75\% of $\text{EER}_{\text{spoof}}$, and by 13.71 and 9.86\% of $\text{EER}_{\text{joint}}$, respectively. Also, the min-tDCF metric, which shows the performance of the biometrics system on a different operating point with respect to the $\text{EER}_{\text{joint}}$ \cite{min_tdcf}, is significatively higher for the proposed ABTN adversarial attacks. As in the white-box scenario, it is worth noticing that the best adversarial attacks do not affect the performance of the ASV system with respect to the baseline system, since the perturbation noise of these attacks is not detected by the ASV system. However, when the hyper-parameter $\epsilon \geq 2.0$, both the PAD and ASV systems are able to detect the perturbations added by the FGSM and PGD methods, and hence, the biometrics system performs even better than the baseline system (denoted by 'No Attack'). However, the proposed ABTN method does not suffer from this issue since it is trained so that the added perturbation noise does not modify the speaker information from the \textit{spoofing} utterance.

\section{Conclusion}
\label{sec:conclusion}

In this work, we studied the robustness of state-of-the-art voice biometrics systems (ASV + PAD) under the presence of adversarial \textit{spoofing} attacks. Moreover, we proposed an adversarial biometrics transformation network (ABTN) for both white-box and black-box scenarios which is able to generate adversarial \textit{spoofing} attacks in order to fool the PAD system without being detected by the ASV system. Experimental results have shown that biometric systems are highly sensitive to adversarial \textit{spoofing} attacks in both logical and physical access scenarios. Moreover, the proposed ABTN system clearly outperforms other popular adversarial attacks such as the FGSM and PGD methods in both white-box and black-box scenarios. In the future, we would like to use the generated adversarial attacks for adversarial training in order to make the biometrics system more robust against these attacks.


\bibliographystyle{IEEEtran}
\bibliography{template}



\end{document}